\documentclass[conference]{IEEEtran}
\usepackage{epsfig,psfrag}
\usepackage{amsmath,amsfonts,amssymb,amsxtra,bm}
\usepackage{graphicx}
\usepackage{tikz}
\usepackage{color}
\usepackage{cite}
\usepackage{setspace}
\usepackage{tikz}
\usepackage{pgfplots}
\usepackage{caption}
\usepackage{notation}

\newcommand{\ist}{\hspace*{.3mm}}
\newcommand{\rmv}{\hspace*{-.3mm}}

\newcommand{\be}{\begin{equation}}
\newcommand{\ee}{\end{equation}}
\newcommand{\iist}{\hspace*{1mm}}
\newcommand{\rrmv}{\hspace*{-1mm}}

\newcommand{\nn}{\nonumber}

\newcommand{\T}{\text{T}}

\providecommand{\rd}{\textcolor{red}}
\newcommand{\va}[1]{#1}

\newcommand{\RFSst}{\Set{X}}
\newcommand{\RFSstR}{\RS{X}}	
\newcommand{\st}{\V{x}}
\newcommand{\stR}{\RV{x}}
\newcommand{\ex}{r}
\newcommand{\sd}{f}

\newcommand{\me}{z}
\newcommand{\meR}{\rv{z}}
\newcommand{\mev}{\V{z}}
\newcommand{\mevR}{\RV{z}}

\newcommand{\ass}{\va{a}}								
\newcommand{\assv}{\V{a}}								
\newcommand{\assR}{\rv{a}}							
\newcommand{\assvR}{\RV{a}}	
\newcommand{\assb}{\va{b}}								
\newcommand{\assvb}{\V{b}}								
\newcommand{\assbR}{\rv{b}}							
\newcommand{\assvbR}{\RV{b}}							
\newcommand{\assw}{\beta}
\newcommand{\su}{p_{\text{S}}}

\newcommand{\PHD}{\lambda}
\newcommand{\card}{\rho}

\allowdisplaybreaks
\makeatletter
\def\blfootnote{\xdef\@thefnmark{}\@footnotetext}
\makeatother

\title{A Scalable Track-Before-Detect Method \\[0.5mm] With Poisson/Multi-Bernoulli Model }
\author{\IEEEauthorblockN{Thomas Kropfreiter\IEEEauthorrefmark{1}, Jason L. Williams\IEEEauthorrefmark{2}, and Florian Meyer\IEEEauthorrefmark{1}\vspace{2mm}}
\IEEEauthorblockA{\IEEEauthorrefmark{1}Scripps Institution of Oceanography and the Department
of Electrical and Computer Engineering, \\University of California
San Diego, La Jolla, CA, USA (e-mail: tkropfreiter@eng.ucsd.edu,\ist flmeyer@ucsd.edu)}
\IEEEauthorblockA{\IEEEauthorrefmark{2}Data61, Commonwealth Scientific and Industrial
Research Organisation, Australia (jason.williams@data61.csiro.au) \vspace{-6mm}}
\thanks{This research was supported by the Office of Naval Research under Grant N00014-21-1-2267.}
}
\IEEEoverridecommandlockouts

\begin{document}

\maketitle

\begin{abstract}
We propose a scalable track-before-detect (TBD) tracking method based on a Poisson/multi-Bernoulli model. 
To limit computational complexity, we approximate the exact multi-Bernoulli mixture posterior probability density function (pdf) by a multi-Bernoulli pdf. Data association based on the sum-product algorithm and recycling of Bernoulli components enable the detection and tracking of low-observable objects with limited computational resources.
Our simulation results demonstrate a significantly improved tracking performance compared to a state-of-the-art TBD method.
\vspace{1mm}
\end{abstract}

\begin{IEEEkeywords}
Multiobject tracking, 
track-before-detect, 
message passing, 
random finite 
\vspace{-2.5mm}
sets.
\end{IEEEkeywords}

\vspace{2mm}
\section{Introduction}
\label{sec:int}
Multiobject tracking aims to estimate the time-dependent number and states of multiple objects based on data provided by one or more sensors. 
The measurements of conventional multiobject tracking methods are produced by a detector that performs thresholding and, possibly, some further preparatory processing of the raw sensor data \cite{Bar11,Cha11,Koc14,Mah07,Wil15,Mey18Proc,Mey17MSBP,Kro16,KroMeyCorCarMenWil:C21}. This preparatory processing step reduces data flow and computational complexity, but results in a loss of tracking information. Thus, the conventional \emph{detect-than-track} approach can lead to a significantly reduced tracking performance, especially in scenarios with low signal-to-noise (SNR) ratio \cite{Ton98,Moy11,Bar85TBD,Dav18TBD,Ris13,Ris19,Vo10TBD,Kim21}.  

In such scenarios, tracking methods that use raw sensor data as measurements, can potentially achieve an improved tracking performance. 
Many existing \emph{track-before-detect} (TBD) methods are based on batch processing using maximum likelihood estimation \cite{Ton98}, the Hough transform \cite{Moy11}, and dynamic programming techniques \cite{Bar85TBD}. 
However, due to their high computational complexity, they are often unsuitable for real-time operation. 
A well-established real-time TBD method is the Histogram Probabilistic Multi-Hypothesis Tracker (H-PMHT) \cite{Dav18TBD}, which is based on the expectation-maximization algorithm. 
However, tuning the parameters of the H-PMHT is known to be difficult \cite{Kim21}. Another class of real-time TBD tracking methods is those derived using random finite sets (RFSs). These methods include the (single) Bernoulli filter \cite{Ris13,Ris19} for single-object tracking and multi-Bernoulli (MB) filters \cite{Vo10TBD,Kim21} for multiobject tracking. Existing MB filters \cite{Vo10TBD,Kim21} for multiobject TBD rely on certain heuristics to introduce Bernoulli components for objects that appear in the scene for the first time.

A state-of-the-art conventional tracking method is the track-oriented marginal multi-Bernoulli/Poisson (TOMB/P) filter \cite{Wil15,Mey18Proc,Kro16}, which models an unknown number of objects and their states by the union of a Poisson RFS and a MB RFS.
The TOMB/P filter relies on a high-dimensional marginalization operation performed by means of message passing \cite{Mey18Proc,Wil14}. A modified version of the TOMB/P filter \cite{Wil12} ``recycles'' Bernoulli components with low existence probability by transferring them to the Poisson RFS instead of pruning them. This recycling step can significantly improve tracking performance in scenarios with reduced sensor performance \cite{Wil12}. 
The TOMB/P filter has also been derived and extended to multiple sensors using the framework of factor graphs and the sum-product algorithm (SPA) \cite{Mey18Proc,Mey17MSBP}. It has been demonstrated that it can outperform existing detect-than-track methods and that it is highly scalable in relevant system parameters \cite{Wil15,Mey18Proc,Mey17MSBP,Kro16}.

In this paper, we propose a scalable TBD method that is based on a Poisson/MB model similar to the one of the TOMB/P filter. We first review the general TBD measurement model and discuss a simplification that is suitable for the case where at any discrete time step, the influence of an object is limited to a single ``bin'' or ``pixel'' of the raw sensor data. Based on this simplified TBD measurement model, we derive an exact update step.
Since the resulting exact posterior probability density function (pdf) is a MB mixture pdf, we approximate it by an MB pdf to limit computational complexity. To guarantee reliable multiobject tracking performance, we recycle Bernoulli components with a low existence probability instead of pruning them.
Finally, we verify the performance of the proposed method numerically and compare it to a state-of-the-art TBD tracking method \cite{Vo10TBD}. 

The main contributions of this paper can be summarized as \vspace{-.3mm} follows\rd{:}
\begin{itemize}
\item We establish a Bayesian Poisson/MB model for the development of scalable TBD methods.
\vspace{.5mm}
\item We derive a scalable Poisson/MB filter for TBD that relies on recycling of Bernoulli components.
\vspace{.5mm}
\item We demonstrate performance improvements of our TBD method compared to a state-of-the-art technique.
\end{itemize}

\emph{Notation:} We will use the following basic notation. Random variables are displayed in sans serif, upright fonts; their realizations in serif, italic fonts. Vectors and matrices are denoted by bold lowercase and uppercase letters, respectively. For example, $\rv{x}$ is a random variable, and $x$ is its realization and $\RV{x}$ is a random vector and $\V{x}$ is its realization. 
Random sets and their realizations are denoted by upright sans serif and calligraphic font, respectively. For example, $\RS{X}$ is a random set and $\Set{X}$ is its realization. We denote pdfs 
by $f(\cdot)$ and probability mass functions (pmfs)
by $p(\cdot)$. $\Set{N}(\V{x}; \V{\mu},\BM{\Sigma})$ denotes the Gaussian pdf  (of random vector $\RV{x}$) with mean $\V{\mu}$ and covariance \vspace{0mm} matrix $\BM{\Sigma}$ and $\Set{R}(x; \sigma)$ denotes the Rayleigh pdf (of scalar random variable $\rv{x}$) with scale parameter $\sigma$. The probability of an event 
is denoted $\mathrm{Pr}\{\cdot\}$. The symbol $\propto$ indicates equality up to a normalization \vspace{-.5mm} factor.


\section{RFS Fundamentals}
\vspace{-.5mm}
\label{sec:fund}

A RFS $\RFSstR$ is a random variable whose realizations $\RFSst$ are finite sets $\big\{\st^{(1)}\rmv,\ldots,\st^{(n)}\big\}$ of vectors $\st^{(i)} \!\in\rmv \mathbb{R}^{n_x}\rmv$.
Both the vectors $\stR^{(i)}$ and their number ${\sf n} \rmv=\rmv |\RFSstR|$ (the cardinality of $\RFSstR$) are random. Thus, $\RFSstR$ consists of a random number ${\sf n}$ of random vectors 
$\stR^{(1)}\rmv,\ldots,\stR^{({\sf n})}\rmv$.
While the conventional Riemann integral is not defined for sets, one can define the set integral 
of a real-valued set function $g(\RFSst)$ \vspace{0mm}as \cite{Mah07}
\begin{align}
&\int g(\RFSst) \ist\ist \delta \RFSst \nn\\
&\hspace{3mm}\triangleq \sum_{n=0}^{\infty} \ist \frac{1}{n!} \ist \int_{\mathbb{R}^{n n_x}} \! g(\{\st^{(1)}\rmv,\ldots,\st^{(n)}\}) 
  \ist\ist \text{d}\st^{(1)}\rmv\cdots \text{d}\st^{(n)}. \label{eq:RFSfund_SetInt} \\[-2mm]
\nn\\[-7.5mm]
\nn
\end{align}
Note that each term of the sum corresponds to one value of the cardinality $n \rmv=\rmv |\RFSst|$.
The statistics of an RFS $\RFSstR$ can be described by its multiobject pdf $f(\RFSst)$.
For any realization $\RFSst = \big\{\st^{(1)}\rmv,\ldots,\st^{(n)}\big\}$, the multiobject pdf $f(\RFSst)$ is\vspace{.5mm} given by
\vspace{0mm} 
\begin{equation}
\label{eq:fund0}
f(\RFSst) \ist=\ist n! \ist \card(n) \ist f_n(\st^{(1)}\rmv,\ldots,\st^{(n)}) \ist.
\vspace{1mm}
\end{equation}
Here, $\card(n) \triangleq \mathrm{Pr}\{|\RFSstR| \rmv=\rmv n\}$, $n \rmv\in \mathbb{N}_0$ is the pmf of ${\sf n} \rmv=\rmv |\RFSstR|$, and $f_n(\st^{(1)}\rmv,\ldots,\st^{(n)})$ is a joint 
pdf of the random vectors $\stR^{(1)}\rmv,\ldots,\stR^{(n)}$ that is invariant to a permutation of its arguments. Note that based on the set integral \eqref{eq:RFSfund_SetInt}, $f(\RFSst)$ integrates to one.

Next, we will review five classes of RFSs \cite{Mah07,Wil15,Mey18Proc} that are relevant for the derivation of the proposed method.
A \textit{Poisson RFS} $\RFSstR$ is characterized by a cardinality that is Poisson distributed with mean 
$\mu$, i.e., $\rho(n) = e^{-\mu}\mu^{n}/n!\ist$, $n \!\in\! \mathbb{N}_0$ and by elements $\stR^{(1)}\rmv,\ldots,\stR^{(n)}$ that are independent and identically distributed (iid) according to the spatial pdf $f(\st)$, i.e, $f_n(\st^{(1)}\rmv,\ldots,\st^{(n)}) = \prod_{i=1}^n f(\st^{(i)})$. Following \eqref{eq:fund0} the 
multiobject pdf now \vspace{.5mm} reads
\[
f^{\text{P}}(\RFSst) \ist=\ist e^{-\int\PHD(\st')\ist\text{d}\st'}\prod_{\st \ist\in \RFSst} \rmv \PHD(\st)
\vspace{-.5mm}
\]
where $\PHD(\st) = \mu f(\st)$ is called the \textit{probability hypothesis density (PHD)} or \textit{intensity function}.

A \textit{Bernoulli RFS} $\RFSstR$ is represented by an existence probability $\ex$ and a spatial pdf $\sd(\st)$. It consists of either none or one element with probability $1-\ex$ and $\ex$, respectively. According to \eqref{eq:fund0}, the multiobject pdf thus\vspace{.5mm} reads
\begin{equation}
\label{eq:fund2}
f(\RFSst) \ist=
\begin{cases} 
1 \!-\rmv \ex , 			 &\RFSst \!=\rmv \emptyset, \\[-.6mm]
\ex \ist \sd(\st) \ist, &\RFSst \!=\! \{\st\}, \\[-.6mm]
0, & \text{otherwise}.
\end{cases}
\vspace{-1.5mm}
\end{equation}

A \textit{MB RFS} $\RFSstR$ is the union of a fixed number $J$ of statistically 
independent Bernoulli RFSs $\RFSstR^{(j)}\rmv$, $j \in \{1,\ldots,J\}$ with multiobject pdfs $f^{(j)}(\RFSst)$ (cf. \eqref{eq:fund2}) described by the
existence probabilities $\ex^{(j)}$ and the spatial pdfs $\sd^{(j)}(\st)$.
The multiobject pdf $f^{\text{MB}}(\RFSst)$ of the MB RFS can be obtained by applying the set convolution \cite{Mah07} to the individual Bernoulli pdfs $f^{(j)}(\RFSst)$. For any realization $\RFSst = \{\st^{(1)}\rmv,\ldots,\st^{(n)}\}$ with $n \leq J$, the multiobject pdf $f^{\text{MB}}(\RFSst)$ can be evaluated\vspace{-1mm} as
\begin{equation}
\label{eq:fund3}
f^{\text{MB}}(\RFSst) = \sum_{\RFSst^{(1)}\uplus\ldots\uplus\ist\RFSst^{(J)} = \RFSst} \prod_{j\ist=\ist 1}^{J}  \ist\ist f^{(j)}(\RFSst^{(j)}) \ist.
\end{equation}
Here, 
\vspace{0.3mm}
$\sum_{\RFSst^{(1)}\uplus\ldots\uplus\RFSst^{(J)} = \RFSst}$ denotes the sum over all disjoint decompositions of $\RFSst$ into sets $\RFSst^{(j)}$, $j\rmv\in\rmv\{1\ldots,J\}$ such that 
$\RFSst^{(1)}\cup\ldots\cup\RFSst^{(J)} = \RFSst$.
For example, for $\RFSst = \big\{\st^{(1)}\rmv\rmv,\st^{(2)}\big\}$ and $J \rmv=\rmv 2$, the multiobject pdf $f^{\text{MB}}(\RFSst)$ \vspace{.5mm} can be evaluated as
\begin{align}
&f^{\text{MB}}\big(\{\st^{(1)}\rmv\rmv,\st^{(2)}\}\big)  \nn\\
&= \ex^{(1)} \ist\ist \ex^{(2)} \ist\ist \Big( f^{(1)}\big(\st^{(1)}\big) \ist f^{(2)}\big(\st^{(2)}\big) +  f^{(1)}\big(\st^{(2)}\big) \ist f^{(2)}\big(\st^{(1)}\big)  \Big).\nn\\[-4.5mm]
\nn
\end{align}
Similarly, for $\RFSst = \big\{\st^{(1)}\big\}$ and $J \rmv=\rmv 3$, the multiobject pdf $f^{\text{MB}}(\RFSst)$ \vspace{.3mm} reads
\begin{align}
f^{\text{MB}}\big(\{\st^{(1)}\}\big)  &= \ex^{(1)} \ist\ist (1-\ex^{(2)} )  (1-\ex^{(3)} )  \ist\ist f^{(1)}\big(\st^{(1)}\big) \nn\\
&\hspace{0mm}+   (1-\ex^{(1)} ) \ist\ist \ex^{(2)} \ist\ist (1-\ex^{(3)} )  \ist\ist f^{(2)}\big(\st^{(1)}\big)  \nn\\
&\hspace{0mm}+  (1-\ex^{(1)} )  \ist\ist (1-\ex^{(2)} ) \ist\ist  \ex^{(3)}  \ist\ist f^{(3)}\big(\st^{(1)}\big).
\nn
\end{align}
Note that for $n \rmv>\rmv J$, we have $f^{\text{MB}}(\RFSst) \rmv=\rmv 0$.

 A \textit{Poisson/MB RFS} $\RFSstR$ is the union of a Poisson RFS and a MB RFS. The pdf of a Poisson/MB RFS can be obtained by applying the set convolution to the pdf of the Poisson RFS and the pdf of the MB RFS. Let $J$ be the number of components of the MB RFS. For any realization $\RFSst = \{\st^{(1)}\rmv,\ldots,\st^{(n)}\}$ with $n \in \mathbb{N}_0$, the multiobject pdf $f^{\text{PMB}}(\RFSst)$\vspace{.3mm} can be evaluated\vspace{-2mm} as
\begin{align}
f^{\text{PMB}}(\RFSst) \hspace{2mm}=\hspace{-3mm} \sum_{\RFSst^{(0)} \uplus\ist \RFSst^{(1)} =\ist\RFSst}\hspace{-3mm} f^{\text{P}}\big(\RFSst^{(0)}\big)\ist f^{\text{MB}}\big(\RFSst^{(1)}\big). \label{eq:fund4}\\[-3.5mm]
\nn
\end{align}
Here, 
$\sum_{\RFSst^{(0)} \uplus\ist \RFSst^{(1)} =\ist\RFSst}$
denotes the sum over all disjoint decompositions of $\RFSst$ into two sets $\RFSst^{(0)}$ and $\RFSst^{(1)}$ such that $\RFSst^{(0)} \cup\ist \RFSst^{(1)} = \RFSst$. 

Finally, a \textit{MB mixture RFS} is a weighted sum of MB RFSs where without loss of generality, we assume that all MB RFSs have the same number of Bernoulli components $J$. For any realization $\RFSst = \{\st^{(1)}\rmv,\ldots,\st^{(n)}\}$ with $n \leq J$, the multiobject pdf $f^{\text{MBM}}(\RFSst)$ can be evaluated\vspace{-.5mm} as
\begin{equation}
\label{eq:fund3_2}
f^{\text{MBM}}(\RFSst) = \sum^{I}_{i\ist=\ist 1} w_i \ist f_i^{\text{MB}}(\RFSst)
\vspace{-1mm}
\end{equation}
where $I$ is the number of different MB pdfs and $\sum^{I}_{i\ist=\ist 1} \rmv w_i = 1$.
Note that in the implementation of our proposed tracking method, the sums in \eqref{eq:fund3}, \eqref{eq:fund4} and \eqref{eq:fund3_2} are never explicitly evaluated.

\vspace{1mm}
\section{System Model}
\label{sec:sys}

In this section, we describe the system model underlying the proposed algorithm.
The multiobject state at time $k$ is represented by an RFS $\RFSstR_{k} = \{\stR_{k}^{(1)},\ldots,\stR_{k}^{(n)}\}$.
The single-object state $\stR_{k}^{(i)}$, $i \rmv\in \rmv\{1,\ldots,n\}$ consists of the object's intensity $\rv{\gamma}_{k}^{(i)}$\rmv, position  $\RV{p}_{k}^{(i)}$, and possibly motion related parameters. 
The measurement $\mevR_k \triangleq [\meR_k^{(1)}\ldots\meR_k^{(M)}]^{\T}\rmv$ at time $k$, consists of $M$ non-negative scalars $\meR_k^{(m)} \geq 0$, $m \rmv\in\rmv\{1,\ldots,M\}$, that represent the measured intensity of the $m$'th ``bin'' or ``pixel''.

\vspace{1mm}
\subsection{State-Transition Model}
\label{sec:stmod}

We use the well-established conventional RFS state-transition model \cite{Mah07,Wil15,Mey18Proc}. At time $k \rmv-\! 1$, an object with state $\stR_{k-1} \!\in\rmv \RFSstR_{k-1}$ either survives or dies with probabilities $\su$ and $1 \!-\rmv \su$, respectively.
If it survives, its new state $\stR_k$ is distributed according to the single-object state transition pdf $f(\st_k|\st_{k-1})$. 

We assume that the states of different objects survive/die and evolve in time independently, i.e., given $\st_{k-1}$,  the single-object state $\stR_k$ is conditionally independent of all the other single-object states $\stR_k'$. 
Thus, conditioned on the multiobject state $\RFSst_{k-1}$, the multiobject state of the survived objects $\RFSstR_k^{\text{S}}$, can be modeled by an MB RFS, i.e., $\RFSstR_k^{\text{S}} \rmv= \bigcup_{\st_{k-1} \in \RFSst_{k-1}} \!\RS{S}_k(\st_{k-1})$. The components $\RS{S}_k(\st_{k-1})$ of this MB RFS are Bernoulli RFSs (cf. \eqref{eq:fund2}) with existence probabilities $\su$ and spatial pdfs
$f(\st_k|\st_{k-1})$. 

Newborn objects are modeled by a Poisson RFS $\RFSstR_{k}^{\text{B}}\rmv$ with mean $\mu_{\text{B}}$, spatial pdf $f_{\text{B}}(\st_k)$ and, hence, PHD $\lambda_{\text{B}}(\st_k) \rmv= \mu_{\text{B}} \ist f_{\text{B}}(\st_k)$. Conditioned on $\RFSst_{k-1}$, surviving objects $\RFSstR_k^{\text{S}}$ are assumed independent of the newborn objects $\RFSstR_k^{\text{B}}$. Thus, for $\RFSst_{k-1}$ fixed, the overall multiobject state at time $k$,  is given \vspace{-1mm}by 
\[
\RFSstR_k = \RFSstR_k^{\text{S}} \cup \RFSstR_k^{\text{B}} 
  = \Bigg( \bigcup_{\st_{k-1} \in\ist \RFSst_{k-1}} \!\!\!\RS{S}_k\big(\st_{k-1}\big) \rmv \Bigg) \cup \RFSstR_k^{\text{B}}\ist . 
	\]
This model defines the state-transition pdf $f(\RFSst_k|\RFSst_{k-1})$, which can be calculated explicitly via set convolution \cite{Mah07}.

\subsection{General Measurement Model}
\label{sec:measmod}
\vspace{0.5mm}




We consider a general superpositional\footnote{The considered superpositional intensity model typically applies to sonar and radar tracking applications. However, it can be easily altered to an occlusion model used in image tracking applications.} intensity model \cite{Kim21} where the influence of object $\stR_k^{(i)}$, $i \rmv\in\rmv \{1,\ldots,n\}$ to the intensity measurement for pixel $\meR_k^{(m)}$, $m \rmv\in\rmv \{1,\ldots,M\}$ is described by an arbitrary point spread function (PSF) $d^{(m)}(\st_k^{(i)})$. The total influence of the multiobject state $\RFSstR_k$ on measurement $\meR_k^{(m)}$, $m \rmv\in\rmv \{1,\ldots,M\}$ is modeled as the sum of all individual PSFs, i.e.,
\vspace{-0.2mm}
\begin{equation}
\label{eq:gen1}
D^{(m)}(\RFSst_k) \ist=\ist \sum_{i\ist=\ist 1}^{n}\iist d^{(m)}(\st_k^{(i)})\ist.
\end{equation}   
This total influence $D^{(m)}(\RFSst_k)$ is a sufficient statistic with respect to measurement $\meR_k^{(m)}$, $m \rmv\in\rmv \{1,\ldots,M\}$. 

For example, the Swerling 1 model \cite{Ris19} often used in radar applications, models all contributions of objects and background noise as statistically independent and circularly symmetric Gaussian random variables. The measurement $\meR_k^{(m)}$, $m \rmv\in\rmv \{1,\ldots,M\}$ is the magnitude of the sum of all contributions. Following the Swerling 1 model, the general likelihood function of measurement $\meR_k^{(m)}$, $m \rmv\in\rmv \{1,\ldots,M\}$ is thus given \vspace{0mm} by the Rayleigh distribution
\[
f_{\mathrm{g}}\big(\me^{(m)}_k|\RFSst_k\big) = \Set{R}\Big(\me_k^{(m)}; \sqrt{D^{(m)}(\RFSst_k) + \sigma^{2}_{\text{n}}}\ist\Big)
\vspace{0mm}
\]
where $\sigma^{2}_{\text{n}}$ is the variance of the background noise.


Conditioned on $\RFSst_k$, all measurements are assumed statistically independent. The general joint likelihood function can thus be obtained as
\vspace{-1.5mm}
\begin{align}
f_{\mathrm{g}}(\mev_k|\RFSst_k) \ist&=\ist \prod_{m \ist=\ist 1}^{M} f_{\mathrm{g}}\big(\me^{(m)}_k|\RFSst_k\big)\ist. \nn \\[-5.5mm]
\nn
\end{align} 
Note that according to this general intensity model, every object can potentially contribute to every\vspace{-2mm} measurement.

\vspace{0.5mm}
\subsection{Considered Simplified Measurement Model}
\label{sec:con_measmod}
\vspace{-1mm}

As an approximation for simplified inference, we consider a model where (i) every object contributes to exactly one distinct measurement $\meR_k^{(m)}\rmv\rmv$, $m \rmv\in \{1,\ldots,M\}$, and (ii) the probability that the object with state $\stR_k^{(i)}$ contributes to measurement $\meR_k^{(m)}$ is proportional to the value of the PSF $d^{(m)}(\st_k^{(i)})$. This simplified model is suitable for scenarios where the PSF of each object is highly concentrated around a single measurement, and objects are unlikely to 
fall 
in 
the same pixel.

To describe the associations of objects and measurements at time $k$, we introduce the vector $\assvR_k \rmv\triangleq\rmv [\assR_k^{(1)},\ldots,\assR_k^{(n)}]^{\T}$ with elements $\assR_k^{(i)}\rmv \in\rmv \{1,\ldots,M\}$. 
Here, $\assR_k^{(i)} = m \in \{1,\ldots,M\}$ indicates that the object with state $\stR_k^{(i)}$, $i \rmv\in\rmv\{1,\ldots,n\}$ only contributes to measurement $\meR_k^{(m)}\rmv$. 
An object-measurement association vector $\assv_k$ is admissible if each measurement involves contributions of at most one object. All admissible associations form the association alphabet $\mathcal{A}_{n,M}$. 
For a fixed association vector $\assv_k \rmv\in\rmv \mathcal{A}_{n,M}$, the total influence of the multiobject state $\RFSstR_k$ on measurement $\meR_k^{(m)}\rmv$, $m \rmv\in\rmv \{1,\ldots,M\}$ is thus given by \vspace{-0.8mm} (cf.~\eqref{eq:gen1})
\vspace{-0.5mm}
\[
\tilde{D}^{(m)}(\RFSst_k) \ist=\ist 
\begin{cases}
d^{(m)}(\st_k^{(i)})\ist, & \exists i\iist \text{such that}\iist \ass_k^{(i)} = m\ist, \\[1.2mm]
0\ist, & \text{otherwise} \ist.
\end{cases}
\]
Hence, if an object $\stR_k^{(i)}$ is associated to measurement $\ass_k^{(i)} \rmv= m$, then the total influence of the multiobject state $\RFSstR_k$ on measurement $\ass_k^{(i)} \rmv=\rmv m$ is $d^{(m)}(\st_k^{(i)})$. If no object is associated to measurement $m$, then there is no influence of the multiobject state $\RFSstR_k$ on measurement $m$. 

As we show in Appendix \ref{sec:App_0}, the joint likelihood function of $\meR_k$ given $\RFSst_k \rmv= \{\st_k^{(1)},\ldots,\st_k^{(n)}\}$ is given by
\vspace{-1mm}
\begin{align}
f(\mev_k|\RFSst_k) &\propto \sum_{\assv_k \in\ist \mathcal{A}_{n,M}} \bigg( \prod_{i\ist=\ist1}^{n} \ist\ist d^{(\ass_k^{(i)})}(\st_k^{(i)}) f_1\big(\me_k^{(\ass_k^{(i)})}\big|\st_k^{(i)}\big) \bigg) \nn\\ 
 &\hspace{15mm}\times \rmv\rmv\rmv\hspace{-.35mm} \prod_{m \ist\in\ist \mathcal{M}_{\V{a}_k}}\rmv\rmv\rmv\rmv\rmv  f_0\big(\me_k^{(m)}\big) \label{eq:meas2_4}
\end{align}	
for $n \rmv\leq\rmv M$ and according to $f(\mev_k|\RFSst_k) = 0$ for $n \rmv>\rmv M$, respectively.
Here, $\mathcal{M}_{\V{a}_k} = \big\{1,\ldots,M\big\} \setminus \big\{  \ass_k^{(1)}, \dots, \ass_k^{(n)} \big\}$ consists of the indexes of all measurements that are not associated to any object state $\stR_k^{(i)}\rmv\rmv$. In addition, $f_0(\me_k^{(m)})$ and $f_1(\me_k^{(m)}|\st_k^{(i)})$ denote the pdfs of measurement $\me_k^{(m)}\rmv, m \rmv\in\{1,\ldots,M\}$ for the cases where it is associated to none or one object $i \rmv\in\rmv\{1,\ldots,n\}$, respectively. For example, following the Swerling 1 model, the pdfs $f_1(\me_k^{(m)}|\st_k^{(i)})$ and $f_0(\me_k^{(m)})$  are given\vspace{1mm} by
\begin{align}
f_1\big(\me_k^{(m)}\big|\st_k^{(i)}\big) &\triangleq \Set{R}\Big(\me^{(m)}_k; \sqrt{d^{(m)}(\st^{(i)}_k) + \sigma^{2}_{\text{n}}} \ist \Big)
\label{eq:meas1} \\[2mm]
f_0\big(\me_k^{(m)}\big) & \triangleq \Set{R}\big(\me^{(m)}_k; \sigma_{\text{n}}\big)\ist. \label{eq:meas2} \\[-3mm]
\nn
\end{align}

\section{Exact Estimation}
\vspace{-2mm}
\label{sec:upd_ex}

In the Bayesian sequential estimation framework, the statistics of the state $\RFSstR_k$ at time $k$, conditioned on all received measurements $\mev_{1:k} \triangleq [\mev_1\ldots\mev_{k}]^{\T}$ up to time $k$, can be described by the posterior pdf $f(\RFSst_k|\mev_{1:k})$. 
This pdf is calculated from the previous posterior pdf $f(\RFSst_{k-1}|\mev_{1:k-1})$ via a prediction and an update step. The prediction step calculates the predicted posterior pdf $f(\RFSst_{k}|\mev_{1:k-1})$ from the previous posterior pdf $f(\RFSst_{k-1}|\mev_{1:k-1})$ based on the state-transition pdf $f(\RFSst_k|\RFSst_{k-1})$ discussed in Section \ref{sec:stmod}. It is identical to the one performed by the TOMB/P filter \cite{Wil15,Mey18Proc,Mey17MSBP,Kro16} and will be skipped. Note that $f(\RFSst_k|\mev_{1:k-1})$ is a Poisson/MB pdf with $J_{k-1}$ Bernoulli components and intensity function $\lambda_{k|k-1}(\st_k)$.

The update step calculates the current posterior pdf $f(\RFSst_{k}|\mev_{1:k})$ from the  predicted posterior pdf $f(\RFSst_k|\mev_{1:k-1})$ based on the likelihood function $f(\mev_k|\RFSst_k)$ in \eqref{eq:meas2_4}. As we show in Appendix \ref{sec:App_A}, the updated posterior pdf $f(\RFSst_k|\mev_{1:k})$ is no longer a Poisson/MB pdf but a MB mixture pdf (cf. \eqref{eq:fund3_2}). 
Contrary to the conventional TOMB/P filter, the updated posterior pdf contains no Poisson part anymore. This is a direct consequence of the fact that the TBD measurement model (cf. Section \ref{sec:con_measmod}) does not consider missed detections. 
However, this is not restrictive for practical scenarios, since objects with very low intensities are also admitted by our model (and thus difficult to distinguish from pure noise).
More precisely, as derived in Appendix \ref{sec:App_A}, the exact updated posterior pdf can be expressed as
\begin{align}
&f(\RFSst_k|\mev_{1:k})  \nn \\[0.5mm] 
&\hspace{1mm} =  \sum_{\assv'_k\ist\in\ist\mathcal{A}'_{J_k,M}} \ist p(\assv'_k)\ist\ist f^{\text{MB}}_{\assv'_k}(\RFSst_k) \ist, \nn \\[-2mm]
&\hspace{1mm} = \hspace{-1.7mm} \sum_{\RFSst_k^{(1)}\uplus\ldots\uplus\RFSst_k^{(J_k)}=\RFSst_k} \ist \sum_{\assv'_k\ist\in\ist\mathcal{A}'_{J_k,M}} \ist p(\assv'_k)\ist \prod_{j \ist=\ist 1}^{J_{k}}  \ist f^{(j,\ass_k^{\prime(j)})}(\RFSst_k^{(j)}) \label{eq:upd3} \\[-6mm]
\nn
\end{align}
where each MB mixture component (i) corresponds to one admissible object-measurement association $\assvR'_k \triangleq [\assR_k^{\prime(1)}\ldots\assR_k^{\prime(J_k)}]^{\T}$, (ii) has $J_k \rmv=\rmv J_{k-1}+M$ Bernoulli components, and (iii) is weighted by the probability $p(\assv'_k)$. 
Note that each Bernoulli component represents an object that potentially exists and that $\assvR'_k$ associates each potential object $j\rmv\in\rmv\{1,\ldots,J_k\}$ to a measurement $m\rmv\in\rmv\{1,\ldots,M\}$.

In contrast to the association vector $\assvR_k$ used for the derivation of the likelihood function in \eqref{eq:meas2_4}, here $\assvR'_k$ accounts also for the possible non-existence of objects.
More precisely, $\assvR'_k$ has entries $\assR_k^{\prime(j)} \rmv\in \{0,1,\ldots,M\}$ for $j \rmv\in\rmv \{1,\ldots,J_{k-1}\}$ and entries $\assR_k^{\prime(j)} \rmv\in\rmv \{0,1\}$ for $j \rmv\in\rmv \{J_{k-1}+1,\ldots,J_{k-1}+M\}$.
Here, for $j \rmv\in\rmv \{1,\ldots,J_{k-1}\}$, $\assR_k^{\prime(j)} =\ist 0$ indicates that an object with state $\stR_k^{(j)}$ does not exist and $\assR_k^{\prime(j)} \rmv= m\rmv\in\rmv \{1,\ldots,M\}$ indicates that it does exist and contribute to measurement $\meR_k^{(m)}$. Furthermore, for $j \rmv\in\rmv \{J_{k-1}+1,\ldots,J_{k-1}+M\}$, $\assR_k^{\prime(j)} = 1$ indicates that a new object with state $\stR_k^{(j)}$ contributes to measurement $m \rmv=\rmv j - J_{k-1}$ and $\assR_k^{\prime(j)} = 0$ that no new object contributes to measurement $m \rmv=\rmv j - J_{k-1}$. All admissible association vectors $\assv'_k$ form the set $\mathcal{A}'_{J_k,M}$. The association pmf $p(\assv'_k)$ is given by
\begin{equation}
\label{eq:upd4}
p(\assv'_k) \propto \prod_{j\ist=\ist 1}^{J_k}\ist \assw_k^{(j,\ass_k^{\prime(j)})}
\vspace{.8mm}
\end{equation}
for $\assv'_k \in \mathcal{A}'_{J_k,M}$ and by $p(\assv'_k) \rmv= 0$ for $\assv'_k \rmv\notin\rmv \mathcal{A}'_{J_k,M}$.
In the following, we provide expressions for the existence probabilities $\ex_k^{(j,m)}$, the spatial pdfs $\sd^{(j,m)}(\st_k)$, and the association weights $\assw_k^{(j,m)}$ used in \eqref{eq:upd3} and \eqref{eq:upd4}.

For each potential object $j \rmv\in\{1,\ldots,J_{k-1}\}$ and measurement $m \in \{1,\ldots,M\}$, we have
\vspace{1mm}
\begin{align}
\ex_k^{(j,m)} & = 1 \label{eq:upd_para1_2} \\[1.5mm]
\sd^{(j,m)}(\st_k) & = \frac{d^{(m)}(\st_k)\ist f_1(\me_k^{(m)}|\st_k) \ist \sd_{k|k-1}^{(j)}(\st_k)}{c_k^{(j,m)}} \label{eq:upd_para1_3}  \\[1.5mm]
\assw_k^{(j,m)} &= \ex_{k|k-1}^{(j)}\ist c_k^{(j,m)} \label{eq:upd_para1_1}  \\[-2.7mm]
\nn
\end{align} 
where we introduced the normalization constant $c_k^{(j,m)} \triangleq \int\rmv d^{(m)}(\st_k)\ist f_1(\me_k^{(m)}|\st_k) \ist \sd_{k|k-1}^{(j)}(\st_k)\ist \text{d}\st_k\ist$.
Here, \eqref{eq:upd_para1_2} indicates that the object modeled by Bernoulli component $j$ exists and contributes to measurement $m$. The state of this object is distributed according to \eqref{eq:upd_para1_3} and the likelihood of this event is characterized by \eqref{eq:upd_para1_1}. 
Furthermore, for $m = 0$ we have $\ex_k^{(j,0)} \rmv=\rmv 0$, $\sd^{(j,0)}(\st_k)$ not defined, and 
\vspace{2mm}
\begin{equation}
\label{eq:upd_para2_1}
\assw_k^{(j,0)} = 1 - \ex_{k|k-1}^{(j)}
\end{equation} 
Here, $\ex_k^{(j,0)} \rmv=\rmv 0$ indicates that the object modeled by Bernoulli component $j$ does not exist. The likelihood of this event is given by \eqref{eq:upd_para2_1}.

For each new potential object $j \rmv=\rmv J_{k-1}+m$, $m \in \{1,\dots,$ $M\}$, we get
\vspace{-1.5mm}
\begin{align}
\ex_k^{(j,1)} & = \frac{c_k^{(j)}}{f_0(\me_k^{(m)}) + c_k^{(j)}} \label{eq:upd_para3_2} \\[2mm]
\sd^{(j,1)}(\st_k) & = \frac{d^{(m)}(\st_k)\ist f_1(\me_k^{(m)}|\st_k)\ist \lambda_{k|k-1}(\st_k)}{c_k^{(j)}} \label{eq:upd_para3_3}  \\[2mm]
\assw_k^{(j,1)} &= f_0(\me_k^{(m)}) + c_k^{(j)} \label{eq:upd_para3_1}  \\[-5mm]
\nn
\end{align}
with normalization constant $c_k^{(j)} \triangleq \int d^{(m)}(\st_k) f_1(\me_k^{(m)}|\st_k)$ $\lambda_{k|k-1}(\st_k)\ist \text{d}\st_k$.
Here, \eqref{eq:upd_para3_2} is the probability that there is a new object that contributes to measurement $m$ (assuming that no other existing object contributed to measurement $m$). The state of this object is distributed according to \eqref{eq:upd_para3_3} and the likelihood of this event is given by \eqref{eq:upd_para3_1}.  
Finally, we\vspace{-.2mm} have $\assw_k^{(j,0)} = 1$, $\sd^{(j,0)}(\st_k)$ not defined, and $\ex_k^{(j,0)} = 0$.

\section{Approximate Update Step}
\vspace{-.2mm}

In the following, we will describe the approximate update step used by our TBD method. This update step approximates the exact MB mixture posterior pdf in \eqref{eq:upd3} by a Poisson/MB pdf. This limits computational complexity and, thus, enables real-time tracking of multiple low-observable objects\vspace{-.7mm}.

\subsection{MB Approximation}
\vspace{-.5mm}

First, we approximate the exact MB mixture posterior pdf in \eqref{eq:upd3} by an MB pdf.
This approximation is based on approximating the association pmf $p(\assv'_k)$ in \eqref{eq:upd4} by the product of its marginals. 
We first extend the association alphabet $\mathcal{A}'_{J_k,M}$ in \eqref{eq:upd4} to $\bar{\mathcal{A}}'_{J_k,M} \triangleq \{0,1,\ldots,M\}^{J_{k-1}}\times\{0,1\}^{M}$. 
Note that $\bar{\mathcal{A}}'_{J_k,M}$ now also contains inadmissible associations.
This does not affect $p(\assv'_k)$ since by definition $p(\assv'_k) = 0$ for $\bar{\mathcal{A}}'_{J_k,M} \setminus \mathcal{A}'_{J_k,M}$. 
Next, we approximate $p(\assv'_k)$ according to 
\vspace{0mm}
\begin{equation}
\label{eq:appox1}
p(\assv'_k) \approx \prod_{j = 1}^{J_k}\ist p\big(\ass_k^{\prime(j)}\big) \ist, \quad \assv'_k \rmv\in\rmv \bar{\mathcal{A}}'_{J_k,M}
\end{equation}
with 
\vspace{1mm}
\[
p\big(\ass_k^{\prime(j)}\big) =\rmv \sum_{\sim\assv_k^{\prime (j)}} \ist p(\assv'_k) \ist.
\vspace{-1mm}
\]	
Here, $\sim \rmv \assv_k^{\prime (j)}$ denotes the vector of all $a_k^{\prime(j')}$ with $j' \in \{1,\ldots,$ $J_k\} \rmv\setminus\rmv j$.
Note that a fast and scalable calculation of $p\big(\ass_k^{\prime(j)}\big)$ is enabled by the SPA \cite{Wil14,Wil15,Mey18Proc}.

Next, we insert \eqref{eq:appox1} into \eqref{eq:upd3}, which yields the approximate posterior pdf $\tilde{f}(\RFSst_k|\mev_{1:k})$ given\vspace{1mm} by 
\begin{align}
&\tilde{f}(\RFSst_k|\mev_{1:k})  \nn \\[-0.5mm] 
& = \sum_{\RFSst_k^{(1)}\uplus\ldots\uplus\ist\RFSst_k^{(J_k)}=\ist\RFSst_k} \ist \sum_{\assv'_k\ist\in\ist\bar{\mathcal{A}}'_{J_k,M}} \ist \prod_{j \ist=\ist 1}^{J_k}\ist\ist p\big(\ass_k^{\prime(j)}\big) \ist f^{(j,\ass_k^{(j)})}(\RFSst_k^{(j)}) \ist. \nn\\[-2.7mm]
\label{eq:approx2} 
\end{align} 
By further using the identity 
$\sum_{\assv'_k \in \bar{\mathcal{A}}'_{J_k,M}} \! \prod_{j\ist=\ist 1}^{J_k} p(\ass_k^{\prime(j)}) = \big(\prod_{j\ist=\ist 1}^{J_{k-1}} \sum_{\ass_k^{\prime(j)}=\ist 0}^{M} p(\ass_k^{\prime(j)})\big) \prod_{j\ist=\ist J_{k-1}+1}^{J_{k}} \sum_{\ass_k^{\prime(j)}=\ist 0}^{1} p(\ass_k^{\prime(j)})$,
\vspace{0.2mm}
we 
can now rewrite \eqref{eq:approx2} as
\vspace{-0.5mm}
\begin{align}
&\tilde{f}(\RFSst_k|\mev_{1:k}) \ist \nn \\[-1mm] 
&=\rmv\sum_{\RFSst_k^{(1)}\uplus\ldots\uplus\ist\RFSst_k^{(J_k)}=\ist\RFSst_k} \prod_{j \ist=\ist 1}^{J_{k-1}} \sum_{\ass_k^{\prime(j)}=\ist 0}^{M} \ist p(\ass_k^{\prime(j)})\ist  f^{(j,\ass_k^{(j)})}(\RFSst_k^{(j)}) \nn \\
&\times \prod_{j \ist=\ist J_{k-1}+1}^{J_{k}}\ist \sum_{\ass_k^{\prime(j)}=\ist 0}^{1} \ist p(\ass_k^{\prime(j)})\ist  f^{(j,\ass_k^{(j)})}(\RFSst_k^{(j)}).  \label{eq:approx3}
\end{align} 
By comparing \eqref{eq:approx3} with \eqref{eq:fund3}, it can be seen that this approximate posterior pdf $\tilde{f}(\RFSst_k|\mev_{1:k})$ is now a MB pdf, i.e., it can be rewritten\vspace{-2.2mm} as
\begin{equation}
\label{eq:approx3_2}
\tilde{f}(\RFSst_k|\mev_{1:k}) \ist\ist = \rmv\rmv\rmv\rmv \sum_{\RFSst_k^{(1)}\uplus\ldots\uplus\ist\RFSst_k^{(J_k)}=\ist\RFSst_k} \prod_{j \ist=\ist 1}^{J_{k}} \ist f^{(j)}(\RFSst_k^{(j)})
\end{equation}
where the existence probabilities and spatial pdfs of the Bernoulli pdfs $f^{(j)}(\RFSst_k^{(j)})$ for $j \in \{1,\ldots,J_{k-1}\}$ are given by
\vspace{-2mm}
\begin{align}
\ex_k^{(j)} &=  \sum_{\ass_k^{\prime(j)} =\ist 1}^{M} p(\ass_k^{\prime(j)}) \label{eq:approx4} \\[0.5mm]
\sd^{(j)}(\st_k) &= \frac{1}{\ex_k^{(j)}} \rmv\rmv \sum_{\ass_k^{\prime(j)} =\ist 1}^{M} p(\ass_k^{\prime(j)})\ist \sd^{(j,\ass_k^{\prime(j)})}(\st_k)\label{eq:approx5} \\[-4.5mm]
\nn
\end{align}
and for $j \rmv\in\rmv \{J_{k-1}+1,\ldots,J_{k}\}$ by
\vspace{1.2mm}
\begin{align}
\ex_k^{(j)} &=  p(\ass_k^{\prime(j)} = 1)\iist \ex_k^{(j,1)} \label{eq:approx6} \\[3mm]
\sd^{(j)}(\st_k) &= \ist \sd^{(j,1)}(\st_k)\ist. \label{eq:approx7} \\[-5mm]
\nn
\end{align}
Note that the idea of approximating a MB mixture pdf by an MB pdf was also used in the derivation 
of the TOMB/P filter \cite{Wil15} and the SPA-based labeled MB filter\vspace{-2.5mm} \cite{Kro19LMB}.
  
\subsection{Recycling of MB Components}
\vspace{-1mm}

The approximate posterior pdf in \eqref{eq:approx3_2} consists of $J_k = J_{k-1}+M$ Bernoulli components, i.e., the number of Bernoulli components increases by $M$ at each time step $k$. This is because newborn objects, modeled by the intensity function $\lambda_{\text{B}}(\st_k)$,  may appear in the scene and potentially contribute to all of the $M$ measurements. However, many Bernoulli components typically have a very low existence probability and are thus unlikely to represent an existing object.

In order for real-time tracking to remain feasible, the number of Bernoulli components has to be limited. Contrary to most multiobject tracking methods (see e.g. \cite{Bar11,Cha11,Koc14,Mah07,Mey18Proc,Mey17MSBP,Kro16} ), rather than pruning components with an existence probability below a fixed threshold and potentially discarding valuable tracking information, we employ the concept of recycling \cite{Wil12}. Here, Bernoulli components $j \rmv\in\rmv \mathcal{J}_k^{\text{R}} \rmv\subseteq\rmv \{1,\ldots,J_k\}$ with an existence probability $\ex_k^{(j)}$ below a predefined threshold $\eta_{\text{R}}$ are ``transferred'' to the Poisson part of the posterior pdf by means of moment matching \cite{Wil12}. This yields the approximated posterior\vspace{-1.3mm} PHD 
\begin{equation}
\label{eq:approx8}
\PHD(\st_k) = \sum_{j\in \mathcal{J}_k^{\text{R}}} \ex_k^{(j)}\ist \sd^{(j)}(\st_k) \nn
\vspace{-1.2mm}
\end{equation}
where $\ex_k^{(j)}$ is given by \eqref{eq:approx4} or \eqref{eq:approx6} and $\sd^{(j)}(\st_k)$ by \eqref{eq:approx5} or \eqref{eq:approx7}, respectively.  
After applying this recycling step, the approximate posterior pdf is again a Poisson/MB pdf. In particular, the Poisson part is represented by the approximate posterior PHD $\PHD(\st_k)$ in \eqref{eq:approx8} and the MB part by the existence probabilities $\ex_k^{(j)}$ and spatial pdfs $\sd^{(j)}(\st_k)$, $j \rmv\in\rmv \{1,\ldots,J_k\}\setminus\mathcal{J}_k^{\text{R}}$ in \eqref{eq:approx4}--\eqref{eq:approx7}.

\section{Numerical Study}
\label{sec:num}

We consider a two-dimensional (2D) simulation scenario with a region of interest (ROI) of $[\text{0}\text{m},\text{64}\text{m}] \times [\text{0}\text{m},\text{64}\text{m}]$. 
We simulated 10 objects during 200 time steps. 
The object states consist of 2D position, 2D velocity, and the object's intensity, i.e., 
$\stR_k \!=\rmv [\RV{p}_{k} \,\, \rv{\gamma}_k]^{\T}\rmv$ with $\RV{p}_{k} = [\rv{p}_{k,1} \,\, \rv{p}_{k,2} \,\, \dot{\rv{p}}_{k,1} \,\, \dot{\rv{p}}_{k,2}]^{\T}$. The kinematic part of the object's state $\RV{p}_{k}$ evolves according to the nearly constant velocity motion model \cite{Bar02} with iid driving noise distributed according to $\mathcal{N}(\V{\epsilon}_{k,\text{p}};\M{0}_4,10^{-3}\M{I}_4)$, and the object's intensity according to a random walk model with iid driving noise distributed according to $\mathcal{N}(\epsilon_{k,\text{I}};0,10^{-4})$, respectively. 
The objects appear at various times before time step 30 and at randomly chosen positions in the area $[\text{17}\text{m},\text{48}\text{m}] \times [\text{17}\text{m},\text{48}\text{m}]$ at each simulation run, and they disappear at various times after time step 170 or when they leave the ROI.
The object's initial velocity is drawn from
$\mathcal{N}(\V{v}_{k};0,\sigma^2_{\text{v}}\M{I}_2)$ with variance $\sigma^2_{\text{v}} = 10^{-2}\rmv$.
We consider two scenarios. All objects appear with an initial intensity of $\gamma_{\text{I}} \rmv=\rmv 10$ and $\gamma_{\text{I}} \rmv=\rmv 4$ in scenarios 1 and 2, respectively. While the scenario is not deliberately constructed to cause objects to come into close proximity, this occurs randomly, and the behavior of the compared methods is observed.

\begin{figure}
\vspace*{.3mm}
\centering
\footnotesize
\begin{minipage}[H!]{0.3\textwidth}
   \input{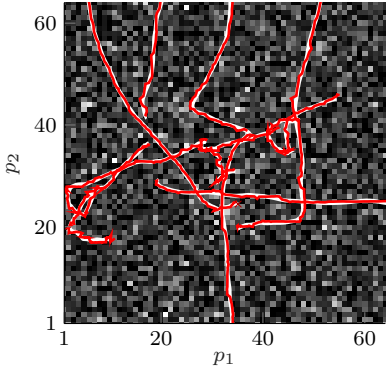}
\end{minipage}
\caption{Example of true trajectories for $\gamma_{\text{I}} \rmv=\rmv 10$ (white lines), as well as of trajectories estimated by the proposed T-TOMB/P filter (red lines)
and measurements acquired at time $k \!=\! 100$.} 
\label{fig:Traj} 
\vspace{-4mm}
\end{figure}

The measurement is an image consisting of $\text{64} \times \text{64}$ cells covering the ROI.
Each cell is a square of $\text{1m}$ side length and has a scalar intensity.
We use the measurement model described in Section \ref{sec:measmod}
and set the PSF to 
\vspace{0.5mm}
\[
d^{(\ass_k^{(i)})}(\st_k^{(i)}) =
\begin{cases}
\gamma_k^{(i)}, & \text{object } \st_k^{(i)} \text{ is in cell } \ass_k^{(i)} \\[1mm]
0, & \text{otherwise}
\end{cases}
\vspace{-2mm}
\]
Hence, $\meR_k^{(m)}$ is distributed according to 
\vspace{-0.3mm}
$\mathcal{R}\big(\me^{(m)}_k; \sqrt{\gamma^{(i)}_k + \sigma^2_{\text{n}}}\big)$ (cf. \eqref{eq:meas1}) if object $\stR_k^{(i)}$ is in cell $m$ and according to $\mathcal{R}\big(\me^{(m)}_k; \sigma_{\text{n}}\big)$ (cf. \eqref{eq:meas2}) if no object is in cell $m$. Here, $\gamma^{(i)}_k$ is the intensity of object $i$ at time $k$ and $\sigma_{\text{n}} \rmv=\rmv 1$ is the standard deviation of the background noise.
If two or more objects $i \rmv\in\rmv\mathcal{I}'_k\rmv \subseteq\rmv\{1,\ldots,I_k\}$ are in the same cell at the same time, we select $i$ by drawing a sample from 
$\rv{i}'\sim \gamma_k^{(i)}$, $i \rmv\in\rmv \mathcal{I}'_k$.
Hence, $\meR_k^{(m)}$ can only have contributions of at most one object (cf. Section \ref{sec:measmod}). 
Thus, the higher the intensity value $\gamma_k^{(i)}$ of object $\stR_k^{(i)}$, the more likely it is to contribute to $\meR_k^{(m)}$. 

\begin{figure*}[t!]
 \vspace*{.3mm}
\centering
\footnotesize

\begin{minipage}[H!]{0.45\textwidth}
\hspace*{-2mm}
  \input{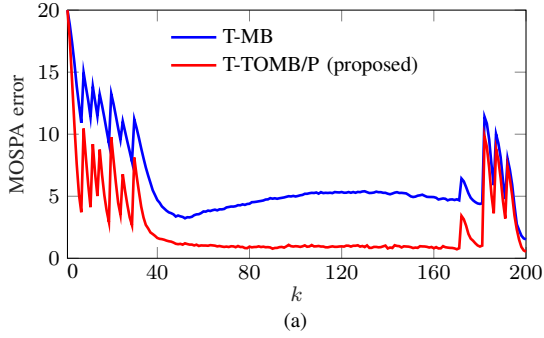}
\end{minipage}
\begin{minipage}[H!]{0.4\textwidth}
\hspace*{1mm}
  \input{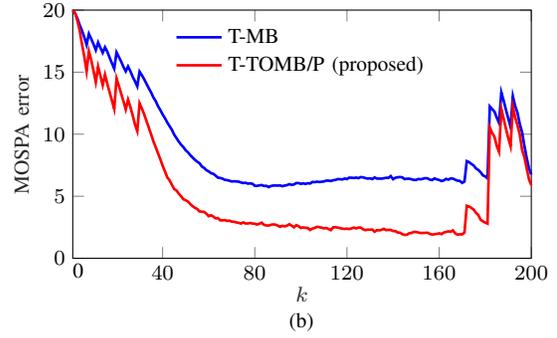}
\end{minipage}

\caption{MOSPA error of T-TOMB/P and T-MB versus time $k$ for (a) $\gamma_{\text{I}} = 10$ and (b) $\gamma_{\text{I}} = 4$.}
\label{fig:results} 
\vspace{-3mm}
\end{figure*}	

We employ a particle implementation of our proposed TBD algorithm, briefly termed T-TOMB/P due to the close relation to the conventional TOMB/P filter \cite{Wil15} for point measurements. 
We compare T-TOMB/P to the TBD-based MB filter proposed in \cite{Vo10TBD}, briefly referred to as T-MB.
Both T-TOMB/P and T-MB represent the spatial pdf of each Bernoulli component by $3,\rmv\rmv000$ particles. T-TOMB/P further represents the posterior PHD by $50,\rmv\rmv000$ particles, the birth PHD by another $50,\rmv\rmv000$, where the resulting $100,\rmv\rmv000$ particles are again reduced to $50,\rmv\rmv000$ after the update step.
More precisely, the birth PHD $\PHD_{\text{B}}(\st_k) \rmv=\rmv \mu_{\text{B}}\ist f_{\text{B}}(\st_k)$ is represented by drawing particles from the pdf $f_{\text{B}}(\st_k) \rmv=\rmv f(p_{k,1} \,\, p_{k,2})\ist f_{\text{v}}\big(\dot{p}_{k,1},\dot{p}_{k,2}\big)\ist f_{\text{I}}(\gamma_{k})$ and by setting $\mu_{\text{B}} \rmv=\rmv 4/64^2$.  
Here, $f(p_{k,1} \,\, p_{k,2})$ is uniform over the ROI, 
$f_{\text{v}}\big(\dot{p}_{k,1},\dot{p}_{k,2}\big)$ is $\mathcal{N}(\dot{p}_{k,1},\dot{p}_{k,2}; \V{0}_2, \sigma^2_{\text{v}}\ist\textbf{I}_2 )$, and
$f_{\text{I}}(\gamma_{k})$ is uniform from $0$ to $\eta_{\text{I}} \rmv=\rmv 30$, respectively.
For the generation of new Bernoulli components of T-MB, we adapt the scheme of \cite{Kim21}.
In fact, T-MB generates a new Bernoulli component for each measurement $\me_{k-1}^{(m)}$ whose intensity value is above the threshold $\eta_{\text{new}} \rmv=\rmv 1.5\ist \sqrt{\gamma_{\text{I}} + \sigma^2_{\text{n}}}$. 
More precisely, the existence probability of each new Bernoulli component is set to $10^{-4}$ and the spatial pdf is represented by drawing particles from the pdf 
$f'_{\text{B}}(\st_k) \propto  \int\rmv f( \st_{k} | \st_{k-1}) \ist 
f\big(\me^{(m)}_{k-1}\ist\big|\ist p_{k-1,1} \,\, p_{k-1,2}\big) \allowbreak\ist  f_{\text{v}}\big(\dot{p}_{k-1,1},\dot{p}_{k-1,2}\big) f_{\text{I}}(\gamma_{k-1}) \ist\text{d}\st_{k-1}$. Here, the function
$f\big(\me_{k-1}^{(m)}\big|p_{k-1,1} \,\, p_{k-1,2}\big)$  is uniform over the cell area of measurement $m$ and $f_{\text{v}}\big(\dot{p}_{k-1,1},\dot{p}_{k-1,2}\big)$ and $f_{\text{I}}(\gamma_{k-1})$ are chosen as in $f_{\text{B}}(\st_k)$. Furthermore, T-TOMB/P recycles Bernoulli components with an existence probability below $\eta_{\text{R}}= 10^{-1}$ and T-MB prunes Bernoulli components with an existence probability below $\eta_{\text{T}}= 10^{-4}$, respectively. T-TOMB/P and T-MB set $\su = 0.999\ist$.

The example shown in Fig. \ref{fig:Traj} suggests excellent detection and estimation performance of the proposed method.
For a quantitative assessment and comparison of the performance of both filters,
we computed the mean Euclidean distance based optimal subpattern assignment (MOSPA) metric \cite{Sch08} 
with cutoff parameter $c \rmv=\rmv 20$, order $p \rmv=\! 2$, and averaged over 1000 simulation Monte Carlo runs. 
Fig. \ref{fig:results} shows the obtained results of T-TOMB/P and T-MB for $\gamma_{\text{I}} \rmv=\rmv 10$ and $\gamma_{\text{I}} \rmv=\rmv 4$. 
It is seen that T-TOMB/P consistently outperforms T-MB. 
This can be attributed to the excellent behavior of the Bernoulli component generation/recycling scheme of T-TOMB/P and the fact that T-TOMB/P considers data association, i.e., a measurement is allowed to be associated to at most one object and an object to exactly one measurement, respectively.   
The performance of T-MB depends greatly on the choice of $\eta_{\text{new}}$. 
A smaller value of $\eta_{\text{new}}$ results in a faster detection of newly appearing objects, 
but in a higher number of false tracks, i.e., Bernoulli components modeling not existing objects, on the other hand.
Another cause for the performance difference can be understood by considering the case where two objects come close to each other. Since T-MB allows a measurement to be associated to more than one object, one of the two objects can be tracked by both tracks and the other object by none after object separation. This is a direct consequence of the fact that T-MB does not consider data \vspace{-3mm} association.

\section{Conclusion}
\vspace{-1mm}
\label{sec:con}

We proposed a scalable track-before-detect (TBD) method for the tracking of low-observable objects that relies on a Poisson/multi-Bernoulli model. To limit the computational complexity, we approximated the exact posterior pdf by a multi-Bernoulli  pdf. For a reliable tracking in real time, a recycling of Bernoulli components is performed.  We demonstrated that with the proposed method, a significant improvement in tracking performance can be achieved compared to a state-of-the-art TBD tracking method. A possible direction for future research is an extension of the proposed method to a more general measurement model by using sum-product algorithms for data association with extended\vspace{-1.5mm} objects \cite{MeyWin:J20,MeyWil:J21}.

\appendices
\renewcommand*\thesubsectiondis{\thesection.\arabic{subsection}}
\renewcommand*\thesubsection{\thesection.\arabic{subsection}}
\vspace{0mm}
\section{} 
\label{sec:App_0}
\vspace{-1mm}

In the following, we derive \eqref{eq:meas2_4}. Based on assumptions (i) and (ii) in Section \ref{sec:con_measmod}, the joint likelihood function for $\RFSst_k = \{\st_k^{(1)},\ldots,\st_k^{(n)}\}$ can be found as \cite{Gar18PMBM}
\vspace{-.2mm}
\[
f(\mev_k|\RFSst_k) \propto \sum_{\RFSst^{(1)}\uplus\ldots\uplus\RFSst^{(M)} = \ist\RFSst} g(\me_k^{(m)}\rmv,\RFSst_k^{(m)})
\]
where we have\vspace{.5mm} introduced
\begin{equation}
\label{eq:App1}
g(\me_k^{(m)}\rmv,\RFSst_k^{(i)}) = 
\begin{cases}
f_0\big(\me_k^{(m)}\big), & \RFSst_k^{(i)} = \emptyset \\[1.5mm]
d^{(m)}(\st_k^{(i)}) \ist f_1\big(\me_k^{(m)}\big|\st_k^{(i)}\big), & \RFSst_k^{(i)} = \{\st_k^{(i)}\} \\[1.5mm]
\end{cases}
\vspace{0.5mm}
\end{equation}
and $\sum_{\RFSst^{(1)}\uplus\ldots\uplus\RFSst^{(M)} = \RFSst}$ denotes the sum over all disjoint decompositions of $\RFSst$ into sets $\RFSst^{(m)}$, $m\rmv\in\rmv\{1,\ldots,M\}$ such that $\RFSst^{(1)}\cup\ldots\cup\RFSst^{(M)} = \RFSst$.


By further inserting \eqref{eq:App1}, we get
\begin{align}
&f(\mev_k|\RFSst_k) \nn \\[0mm]
&\propto \sum_{\RFSst^{(1)}\uplus\ldots\uplus\RFSst^{(M)} = \ist\RFSst} \rmv\rmv \bigg( \rmv \prod_{m:\ist\RFSst_k^{(m)}=\ist\{\st_k^{(i)}\}}\rrmv d^{(m)}(\st_k^{(i)}) \ist f_1\big(\me_k^{(m)}\big|\st_k^{(i)}\big) \rmv \bigg) \nn \\[1mm]
&\hspace{0mm}\times\prod_{m':\ist\RFSst_k^{(m')}=\ist\emptyset} f_0\big(\me_k^{(m')}\big)\ist. \label{eq:App2}
\end{align}
Expression \eqref{eq:App2} can be reformulated by using the object-measurement association vector $\assvR_k$ introduced in Section \ref{sec:con_measmod}, which yields the final expression for the joint likelihood function\vspace{-4.5mm} in \eqref{eq:meas2_4}. 

\vspace{0mm}
\renewcommand*\thesubsectiondis{\thesection.\arabic{subsection}}
\renewcommand*\thesubsection{\thesection.\arabic{subsection}}
\section{} 
\label{sec:App_A}
\vspace{-1mm}

In this appendix, we derive the exact posterior pdf $f(\RFSst_k|\mev_{1:k})$ in \eqref{eq:upd3}. 
First, we obtain the posterior pdf by applying Bayes theorem \cite{Mah07} according to 
\vspace{1mm}
\begin{equation}
\label{eq:upd2}
f(\RFSst_k|\mev_{1:k}) \propto f(\mev_k|\RFSst_k)\ist f(\RFSst_k|\mev_{1:k-1}) \ist.
\vspace{1mm}
\end{equation}
Next, we reformulate the likelihood function in \eqref{eq:meas2_4} as follows.
We define the measurement-object association vector $\assvbR_k \triangleq [\assbR_k^{(1)},\ldots,\assbR_k^{(M)}]$ with entries $\assbR_k^{(m)} \rmv\in\rmv \{0,1,\ldots,n\}$, which expresses the same information as $\assvR_k$ but in a different form \cite{Mey18Proc,Wil14}. In particular, $\assbR_k^{(m)} = i\rmv\in\rmv\{1,\ldots,n\}$ indicates that object $i$ contributes to measurement $m$ and $\assbR_k = 0$ indicates that no object contributes to measurement $m$.
An association $\assvb_k$ is admissible if any object contributes to exactly one measurement $m$, and at most one object contributes to each measurement. All admissible measurement-object associations form the set $\mathcal{B}_{M,n}$.

The combination of both $\assvR_k$ and $\assvbR_k$ is now used to reformulate \eqref{eq:meas2_4} for each $\RFSst_k \rmv=\rmv \{\st_k^{(1)}\ldots,\st_k^{(n)}\}$
according to
\vspace{0mm}
\begin{align}
&f(\mev_k|\RFSst_k) \nn \\
&\propto \hspace{-1mm}\sum_{\RFSst_{k,0}\uplus\RFSst_{k,1} = \RFSst_k} \sum_{\assv_k\ist\in\ist\mathcal{A}_{n_1,M}}\prod_{i\ist=\ist1}^{n_1} \ist\ist f_1\big(\me_k^{(\ass_k^{(i)})}|\st_{k,1}^{(i)}\big)  \ist\ist d^{(\ass_k^{(i)})}(\st_{k,1}^{(i)})\nn \\[0.8mm]
&\times  \hspace{.5mm}\sum_{\assvb_k \in \mathcal{B}_{\mathcal{M}_{\V{a}_k},n_0}}  \hspace{.6mm} \prod_{m\ist\in\ist \mathcal{M}_{\V{a}_k}} g\big(\me_k^{(m)},\RFSst_{k,0}^{(\assb_k^{(m)})}\big) \label{eq:meas8}
\end{align}
with $n \rmv\leq\rmv M$. Here, 
\vspace{0.4mm}
$\RFSst_k$ is decomposed into arbitrary subsets $\RFSst_{k,0} = \{\st_{k,0}^{(1)},\ldots,\st_{k,0}^{(n_0)}\}$ 
\vspace{-0.4mm}
and $\RFSst_{k,1} = \{\st_{k,1}^{(1)},\ldots,\st_{k,1}^{(n_1)}\}$ with $\RFSst_k = \RFSst_{k,0} \cup \RFSst_{k,1}$ and $n \rmv= n_0 + n_1$.
\vspace{-0.2mm}
Further, $g\big(\me_k^{(m)},\RFSst_{k,0}^{(\assb_k^{(m)})}\big)$ is given by 
\vspace{0.2mm}
\eqref{eq:App1} with $\RFSst_{k,0}^{(\assb_k^{(m)})} \rmv= \emptyset$ for $\assb_k^{(m)} \rmv=\rmv 0$ and $\RFSst_{k,0}^{(\assb_k^{(m)})} \rmv= \{\st_k^{(i)}\}$ for $\assb_k^{(m)} \rmv=\rmv i$ and 	
we introduced $\mathcal{B}_{\mathcal{M}_{\V{a}_k},n_0} = \big\{ \V{b}_k \in \mathcal{B}_{M,n_0} \ist | \ist  b_k^{(m)} = 0, \ist\forall m \notin  \mathcal{M}_{\V{a}_k} \big\}$. 
Recall that $f(\mev_k|\RFSst_k)$ 
\vspace{0.1mm}
in \eqref{eq:meas8} is $0$ for $n \rmv>\rmv M$.

Next, we recap that the predicted posterior pdf is of Poisson/MB form and can be represented by
\begin{align}
&f(\RFSst_{k}|\me_{1:k-1}) \nn \\[-1mm]
& =\sum_{\RFSst_{k,0} \uplus\RFSst_{k}^{(1)}\uplus\ldots\uplus\RFSst_{k}^{(J_{k-1})}\ist  =\ist \RFSst_{k}} \rrmv\rrmv f^{\text{P}}(\RFSst_{k,0})\ist \prod_{j\ist=\ist1}^{J_{k-1}} f^{(j)}(\RFSst_{k,1}^{(j)}) \label{eq:app3_2}
\end{align}
where the Bernoulli pdfs $f^{(j)}(\RFSst_{k})$ are parametrized by the existence probabilities $\ex_{k|k-1}^{(j)}$ and spatial pdfs $\sd_{k|k-1}^{(j)}(\st_k)$ and the Poisson pdf $f^{\text{P}}(\RFSst_{k,0})$ is represented by the posterior PHD $\PHD_{k|k-1}(\st_k)$. 
By plugging \eqref{eq:meas8} and \eqref{eq:app3_2} into \eqref{eq:upd2} and performing certain reformulations, we obtain \cite{Gar18PMBM}
\vspace{0mm}
\begin{align}
&f(\RFSst_k|\mev_{1:k}) \nn \\[1mm]
&\hspace{1mm}\propto  \rrmv\rmv \sum_{\RFSst_{k,0}\uplus\RFSst_{k,1} = \RFSst_k}   \sum_{\bar{\assv}_k \in \bar{\mathcal{A}}_{J_{k-1},M}}  \rrmv\rrmv\rmv h_1(\RFSst_{k,1}^{(1)},\ldots,\RFSst_{k,1}^{(J_{k-1})},\bar{\assv}_k)  \nn \\[0mm]
&\hspace{3.2mm}\times \ist \sum_{\mathcal{B}_{\mathcal{M}_{\V{a}_k}, n_0 }} \rrmv h_0(\RFSst_{k,0}^{(1)},\ldots,\RFSst_{k,0}^{(M)},\assvb_k,\bar{\assv}_k)\ist. \label{eq:app1} \\[-7mm]
\nn
\end{align}
Here the association vector $\bar{\assv}_k \rmv\triangleq\rmv [\bar{\ass}^{(1)}_k,\ldots,\bar{\ass}^{(J_{k-1})}_k]$, $j\in \{1,\ldots,J_{k-1}\}$ has entries $\bar{\ass}^{(j)}_k \rmv\in\rmv \{0,1,\ldots,M\}$ where $\bar{\ass}^{(j)}_k \rmv=\rmv 0$ indicates that an object with state $\stR_k^{(j)}$ does not exist and $\bar{\assR}_k^{(j)} \rmv= m\rmv\in\rmv \{1,\ldots,M\}$ indicates that it does exist and contribute to measurement $\meR_k^{(m)}$. All admissible association vectors $\bar{\assv}_k$ form the\vspace{-3mm} set $\bar{\mathcal{A}}_{J_k,M}$.

In addition, the function $h_1(\RFSst_{k,1},\bar{\assv}_k)$ is given by
\vspace{-1mm}
\begin{equation}
\label{eq:app3}
h_1(\RFSst_{k,1}^{(1)},\ldots,\RFSst_{k,1}^{(J_{k-1})},\bar{\assv}_k) \triangleq \prod_{j\ist=\ist 1}^{J_{k-1}} g_1\big(\me_k^{(\bar{\ass}_{k}^{(j)})}|\RFSst_{k,1}^{(j)}\big)\ist f^{(j)}\big(\RFSst_{k,1}^{(j)}\big).
\vspace{0.5mm}
\end{equation}
Here, $f^{(j)}(\RFSst_{k,1}^{(j)})$ is the Bernoulli pdf in \eqref{eq:app3_2} and $g_1\big(\me_k^{(\bar{\ass}_k^{(j)})}|\RFSst_{k,1}^{(j)}\big)$ is obtained as
\begin{align}
&g_1\big(\me_k^{(\bar{\ass}_k^{(j)})}|\RFSst_{k,1}^{(j)}\big) \nn \\[1.2mm]
&\hspace{1mm}=\begin{cases}
d^{(m)}\big(\st_{k,1}^{(i)})\ist f_1(\me_k^{(m)}|\st_{k,1}^{(i)}\big)\ist, & \bar{\ass}_k^{(j)} = m,\ist \RFSst_{k,1}^{(j)} = \{\st_{k,1}^{(i)}\}\ist \\[1mm]
1\ist, & \bar{\ass}_k^{(j)} = 0,\ist \RFSst_{k,1}^{(j)} = \emptyset \ist \\[1mm]
0\ist, & \text{otherwise.} \ist
\end{cases} \nn
\vspace{0.5mm}
\end{align}
Expression \eqref{eq:app3} is a product of weighted Bernoulli pdfs with parameters as in \eqref{eq:upd_para1_2}--\eqref{eq:upd_para2_1}. Furthermore, the function $h_0(\RFSst_{k,0},\assvb_k,\bar{\assv}_k)$ reads
\vspace{-.5mm}
\begin{align}
&h_0(\RFSst_{k,0}^{(1)},\ldots,\RFSst_{k,0}^{(M)},\assvb_k,\bar{\assv}_k) \nn \\[1mm]
&\hspace{15mm}\triangleq\rmv \prod_{m\ist \in\ist \mathcal{M}_{\bar{\V{a}}_k}} g\big(\me_k^{(m)},\RFSst_{k,0}^{(\assb_k^{(m)})}\big)\ist f^{\text{P}}\big(\RFSst_{k,0}^{(\assb_k^{(m)})}\big) \label{eq:app4}\\[-5.5mm]
\nn
\end{align}
where $g\big(\me_k^{(m)},\RFSst_{k,0}^{(\assb_k^{(m)})}\big)$ is 
\vspace{-0.2mm}
given by \eqref{eq:App1} with $\RFSst_{k,0}^{(\assb_k^{(m)})} \rmv= \emptyset$ for $\assb_k^{(m)} \rmv=\rmv 0$ and $\RFSst_{k,0}^{(\assb_k^{(m)})} \rmv= \{\st_k^{(i)}\}$ for $\assb_k^{(m)} \rmv=\rmv i$, 
\vspace{0.2mm}
and $f^{\text{P}}\big(\RFSst_{k,0}^{(\assb_k^{(m)})}\big)$ is the Poisson pdf in \eqref{eq:app3_2}.
Expression \eqref{eq:app4} is a product of weighted Bernoulli pdfs with the parameters as in \eqref{eq:upd_para3_3}--\eqref{eq:upd_para3_1}.
By using the association vector $\assvR'_k$ as defined in Section \ref{sec:upd_ex}, we can now reformulate \eqref{eq:app1} according to	
\begin{align}
&f(\RFSst_k|\mev_{1:k}) \nn \\[-1mm]
&\hspace{3mm}\propto \rrmv\rmv \sum_{\uplus_{i\ist=\ist 1}^{J_k}\RFSst_k^{(i)} = \RFSst_k}\sum_{\assv'_k \in \mathcal{A}_{J_{k},M}} \prod_{j\ist=\ist 1}^{J_k} \ist \assw_k^{(j,\ass_k^{\prime(j)})}\ist f^{(j,\ass_k^{\prime(j)})}(\RFSst_k^{(j)}) \ist. \nn
\end{align}
Finally, we introduce the association pmf $p(\assv'_k)$ (cf. \eqref{eq:upd4}) and obtain the final expression for the posterior pdf in \eqref{eq:upd3}.

\renewcommand{\baselinestretch}{1.02}
\selectfont
\bibliographystyle{IEEEtran}
\bibliography{references}

\end{document}